\documentclass[onecolumn,english,conference]{IEEEtran}
\pdfoutput=1
\usepackage[T1]{fontenc}
\usepackage[active]{srcltx}
\usepackage{color}
\usepackage{babel}
\usepackage{array}
\usepackage{float}
\usepackage{textcomp}
\usepackage{mathrsfs}
\usepackage{multirow}
\usepackage{amsmath}
\usepackage{amssymb}
\usepackage{stackrel}
\usepackage{graphicx}
\usepackage[unicode=true,
 bookmarks=true,bookmarksnumbered=true,bookmarksopen=true,bookmarksopenlevel=1,
 breaklinks=false,pdfborder={0 0 0},pdfborderstyle={},backref=false,colorlinks=false]
 {hyperref}
\hypersetup{pdftitle={Your Title},
 pdfauthor={Your Name},
 pdfpagelayout=OneColumn, pdfnewwindow=true, pdfstartview=XYZ, plainpages=false}

\makeatletter

\providecommand{\tabularnewline}{\\}
\floatstyle{ruled}
\newfloat{algorithm}{tbp}{loa}
\providecommand{\algorithmname}{Algorithm}
\floatname{algorithm}{\protect\algorithmname}

\IEEEoverridecommandlockouts
\ifCLASSOPTIONcompsoc
\usepackage[caption=false,font=normalsize,labelfont=sf,textfont=sf]{subfig}
\else
\usepackage[caption=false,font=footnotesize]{subfig}
\fi
\usepackage{cite}
\usepackage{amsthm}

\newtheorem{proposition}{\textbf{Proposition}}
\usepackage{algorithm}
\usepackage{amsmath}

\usepackage{cite}
\usepackage{bm}
\usepackage{algorithmic}
\usepackage{algorithm}
\usepackage{graphicx}
\renewcommand{\fnum@figure}{Fig.~\thefigure}
\IEEEoverridecommandlockouts

\makeatother

\begin{document}
\title{Coordinated Intra- and Inter-system Interference Management in Integrated
Satellite Terrestrial Networks}
\author{\IEEEauthorblockN{Ziyue Zhang, Min Sheng, \textit{Senior Member},\textit{
IEEE}, Junyu Liu, \textit{Member, IEEE}, and Jiandong Li, \textit{Fellow,
IEEE} }\\
\IEEEauthorblockA{State Key Laboratory of Integrated Service Networks,
Xidian University, Xi'an, Shaanxi, 710071, China\\
Email: zhangziyue@stu.xidian.edu.cn, msheng@mail.xidian.edu.cn, junyuliu@xidian.edu.cn,
jdli@mail.xidian.edu.cn}\thanks{Ziyue Zhang, Junyu Liu, Min Sheng, and Jiandong Li are with the State
Key Laboratory of Integrated Service Networks, Xidian University,
Xi'an, Shaanxi, 710071, China. (e-mail: zhangziyue@xidian.edu.cn;
junyuliu@xidian.edu.cn; \{msheng, jdli\}@mail.xidian.edu.cn).}\thanks{(\textit{Corresponding author: Junyu Liu.})}}
\maketitle
\begin{abstract}
Leveraging the advantage of satellite and terrestrial networks, the
integrated satellite terrestrial networks (ISTNs) can help to achieve
seamless global access and eliminate the digital divide. However,
the dense deployment and frequent handover of satellites aggravate
intra- and inter-system interference, resulting in a decrease in downlink
sum rate. To address this issue, we propose a coordinated intra- and
inter-system interference management algorithm for ISTN. This algorithm
coordinates multidimensional interference through a joint design of
inter-satellite handover and resource allocation method. On the one
hand, we take inter-system interference between low earth orbit (LEO)
and geostationary orbit (GEO) satellites as a constraint, and reduce
interference to GEO satellite ground stations (GEO-GS) while ensuring
system capacity through inter-satellite handover. On the other hand,
satellite and terrestrial resource allocation schemes are designed
based on the matching idea, and channel gain and interference to other
channels are considered during the matching process to coordinate
co-channel interference. In order to avoid too many unnecessary handovers,
we consider handover scenarios related to service capabilities and
service time to determine the optimal handover target satellite. Numerical
results show that the gap between the results on the system sum rate
obtained by the proposed method and the upper bound is reduced as
the user density increases, and the handover frequency can be significantly
reduced.

$\vphantom{}$

\end{abstract}

\section{Introduction \label{sec:Introduction}}

In the emerging 6th generation (6G) wireless communication networks,
the integrated satellite terrestrial networks (ISTNs) can provide
expanded coverage and seamless connectivity \cite{10045716,9693471,8558507}.
ISTNs offer wide signal coverage for ground users (GUs) in remote
areas and high-quality services for hotspots while ensuring communication
quality \cite{9312798,0Space,8643836}. However, the co-channel interference
in the ISTN and the collinear interference between geostationary (GEO)
and low earth orbit (LEO) satellites degrades the downlink rate of
ISTNs. With the increasing number of deployed LEO satellites, intra-system
co-channel and inter-system collinear interference is further aggravated.
Moreover, the fast movement of LEO satellites makes the inter-satellite
handover occur frequently, which makes the interference management
problem more complex. Therefore, an effective handover strategy is
required to mitigate interference and optimize resource allocation.
Considering the synergy between intra- and inter-system interference,
it is necessary to manage the intra- and inter-system interference
in ISTN through the inter-satellite handover and resource allocation.

\subsection{\textit{Related work}}

In ISTN, intra-system interference management has been extensively
investigated. The design of the resource allocation method for power
and spectrum is essential for interference management in ISTN \cite{2011Dynamic,2006Capacity,2016Performance,9693912,7360145}.
To reduce intra-system interference, a dynamic power allocation algorithm
suitable for satellites operating in the Ka-band was proposed in \cite{2011Dynamic}.
However, the influence of cloud and rain decline was not considered,
resulting in an inaccurate description of the interference. Furthermore,
an optimal bandwidth constraint was proposed in \cite{2006Capacity},
which reduced intra- and inter-beam interference by optimizing frequency
bandwidth and power allocIn the emerging 6th generation (6G) wireless
communication networks, the integrated satellite terrestrial networks
(ISTNs) can provide expanded coverage and seamless connectivity \cite{10045716,9693471}.
ISTNs offer wide signal coverage for ground users (GUs) in remote
areas and high-quality services for hotspots while ensuring communication
quality \cite{9312798,0Space}. However, the co-channel interference
in the ISTN and the collinear interference between geostationary (GEO)
and low earth orbit (LEO) satellites degrades the downlink rate of
ISTNs. With the increasing number of deployed LEO satellites, intra-system
co-channel and inter-system collinear interference is further aggravated.
Moreover, the fast movement of LEO satellites makes the inter-satellite
handover occur frequently, which makes the interference management
problem more complex. ation. Then, the scenario of co-frequency interference
in ISTN was first analyzed in \cite{2016Performance}. Simulation
results illustrate that dynamic resource allocation could effectively
reduce intra-system interference compared to fixed resource allocation.
A resource allocation scheme with interference cooperation was proposed
in \cite{9693912}, in which the original problem was divided into
three stages to be solved sequentially. Although the above researches
can effectively alleviate intra-system interference, with the intensive
deployment of LEO satellites, the collinear interference between LEO
and GEO satellites needs to be considered.

For inter-system interference in ISTN, power allocation and spatial
isolation can be employed to avoid interference \cite{2014Inline,2016Cognitive,2018Uplink,2018Coexistence,2019Optimal}.
In \cite{2014Inline}, a collinear interference suppression method
for GEO and LEO satellite spectrum sharing was proposed. Collinear
interference is mitigated by optimizing the transmit power of the
satellites. In addition, a cognitive distance power control scheme
was proposed to avoid the interference of LEO satellites to GEO satellites
by controlling the distance between satellite system terminals \cite{2016Cognitive}.
In \cite{2018Uplink}, interference caused by the operation of the
LEO system uplink and downlink on the GEO system was analyzed, and
a method of avoiding interference by setting an exclusion angle was
proposed in \cite{2018Coexistence}, respectively. Based on the above
analysis of interference, a power control technique was proposed to
increase the downlink rate under the GEO quality of service (QoS)
constraint in \cite{2019Optimal}. However, the existing power control
methods compromise the system capacity, and spatial isolation comes
at the cost of communication quality. 

The highly dynamic nature of the network topology results in frequent
handovers between terrestrial terminals and satellites, which also
leads to a decrease in the performance of ISTN \cite{9003618}. Inter-satellite
handover is considered as an effective solution for inter-system interference
and resource allocation in ISTN \cite{2016A,2020User,2020A}. In \cite{2016A},
a weighted bipartite graph based handover strategy was proposed for
the links between the satellites and gateway stations. The channel
gain is used as the handover weight. However, inter-satellite handover
would change the LEO satellites that can be connected by TBSs, which
may affect the distribution of co-channel interference and degrade
the backhaul capacity. Therefore, matching games have been introduced
to solve the problem of co-channel interference management in ISTN
\cite{2021Joint,2021Resource,2022A,2022B,2022Cooperative}. A game
matching based resource allocation method was proposed to improve
the transmission rate in \cite{2021Joint}. In addition, a dynamic
spectrum partitioning strategy was proposed to to mitigate interlayer
interference in \cite{2022B}. Through the matching process, both
interference mitigation and user fairness can be considered, and the
downlink sum rate can be improved.

Despite the extensive literature on intra- and inter-system interference
management in ISTN, most existing studies ignore the combined impact
of intra- and inter-system interference on system capacity. The dense
distribution of LEO satellites leads to aggravated inter-system interference,
which may reduce the system capacity and affect the intensity of intra-system
co-channel interference. For comprehensive management of complex interference
in ISTNs, it is necessary to develop a unified framework to effectively
integrate interference management at intra- and inter-system levels,
which could capture the effect of inter-system interference on system
capacity and facilitate the coordinated management of intra-system
co-channel interference.

\subsection{\textit{Contribution}}

In this paper, a coordinated intra- and inter-system interference
management (CIIM) algorithm is proposed for ISTN. CIIM leverages the
correlation of intra- and inter-system interference to conduct coordination
between the LEO and GEO satellites system, and coordination of intra-system
interference in ISTN through resource allocation. The main contributions
of this paper are summarized as follows.
\begin{itemize}
\item For inter-system collinear interference, an interference management
algorithm based on inter-satellite handover is proposed. The algorithm
enables the handover of LEO satellites through the working threshold
to avoid collinear interference. Since frequent handovers will lead
to a decline in real-time decision making, we introduce a handover
threshold based on the signal strength difference to remove LEO satellites
that cannnot meet the handover conditions. By adjusting the handover
threshold, the trade-off between capacity improvement and handover
frequency can be balanced.
\item To coordinate intra-system co-channel interference, we propose a user
association and resource allocation method based on game matching.
During the matching process, GUs are inclined to select subchannels
with high channel power gain and low co-channel interference. Moreover,
TBSs dynamically adjust power allocation according to user distribution
and channel conditions to reduce interference to other GUs. With the
increasing user density, we show via simulation results that the gap
between the results on the system sum rate obtained by the proposed
method and the upper bound is reduced.
\end{itemize}
$\quad$The remainder of this paper is organized as follows. \textcolor{black}{Firstly,
the system model is introduced in Section II. In Section }III,\textcolor{black}{{}
the intra- and inter-system interference management problem is formulated
and decomposed into two subproblems. Then, the }CIIM\textcolor{black}{{}
scheme is detailed in Section IV. In Section V, the proposed algorithms
are evaluated through simulations. The paper concludes in Section
VI.}

\section{System Model \label{sec:System-Model-And-Problem-Formulation}}

\textcolor{black}{In this section, the ISTN model is briefly introduced.
Then the cached model, terrestrial transmission model and satellite
transmission model are given. The notations and definitions used in
this paper are shown in Tab }\ref{tab:Parameter used}\textcolor{black}{. }

\begin{table}[t]
\centering{}\caption{\label{tab:Parameter used} Notations and Definitions}
\begin{tabular}{|c|c|}
\hline 
\textbf{Notations} & \textbf{Definitions}\tabularnewline
\hline 
\hline 
$\mathcal{M}$, $\mathcal{J}$, $\mathcal{N}$, $\mathcal{L}$  & Sets of TBSs, GUs and LEO satellites.\tabularnewline
\hline 
$M$, $J$, $N$, $L$ & Number of TBSs, GUs and LEO satellites.\tabularnewline
\hline 
$\mathcal{C}$, $\mathcal{K}$ & Sets of subchannels available for TBSs and LEO satellites.\tabularnewline
\hline 
$\mathit{N}_{\textrm{C}}$, $\mathit{N}_{\textrm{K}}$ & Number of subchannels available for TBSs and LEO satellites.\tabularnewline
\hline 
$G_{T}$, $G_{R}$ & Gain of the transmit and the receive antennas.\tabularnewline
\hline 
$p_{leo}$ & The transmit power of LEO satellites.\tabularnewline
\hline 
$\mathit{B}_{\textrm{C}}$, $\mathit{B}_{\textrm{Ka}}$ & The bandwidth of C-band and Ka-band.\tabularnewline
\hline 
\end{tabular}
\end{table}

\subsection{Network Model\label{subsec:Network Model}}

\begin{figure}[t]
\begin{centering}
\includegraphics[width=3.5in]{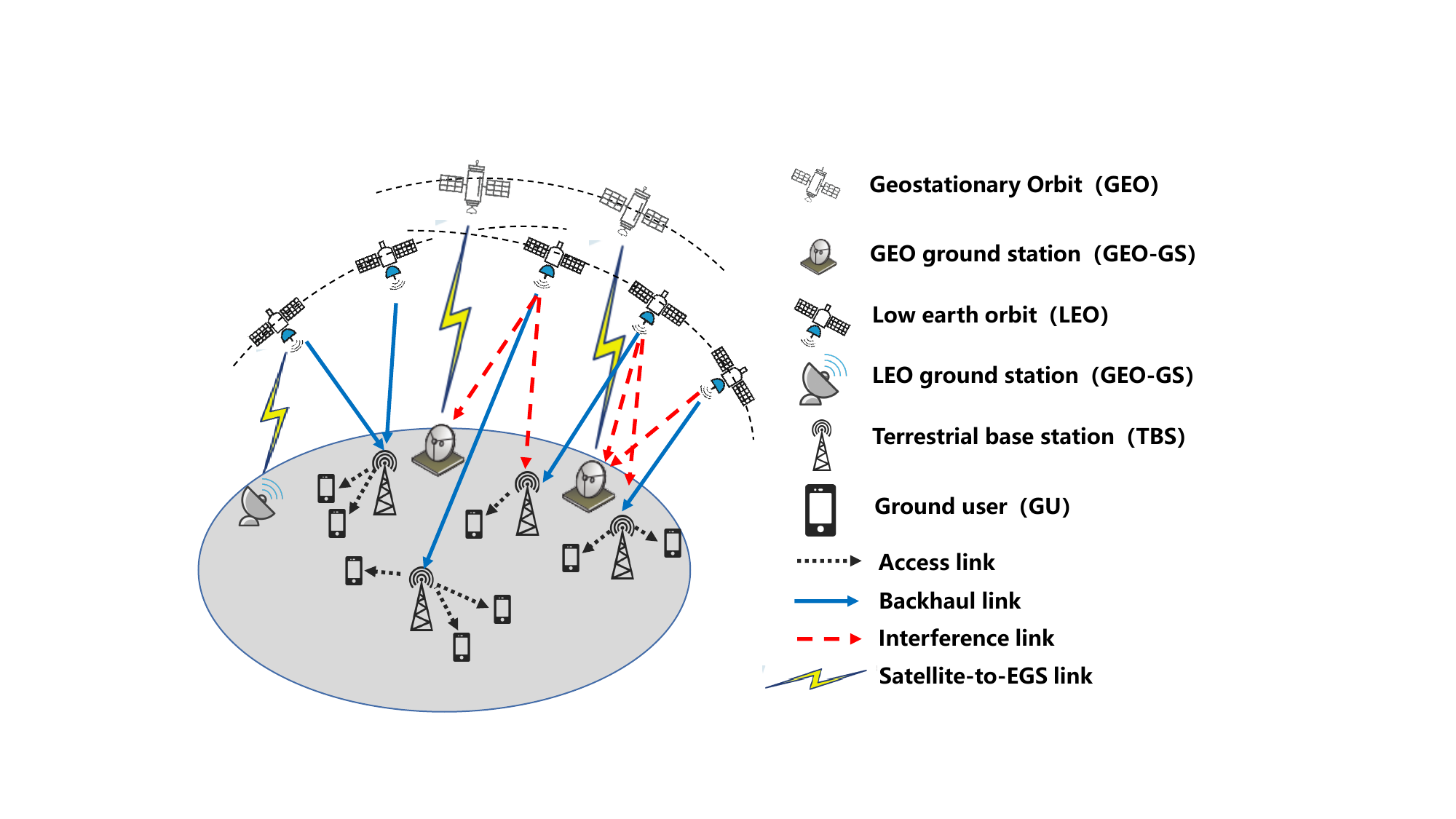}
\par\end{centering}
\centering{}\caption{\label{fig:system model}System model of\textcolor{black}{{} }downlink
ISTNs.$\qquad\qquad\qquad\qquad\quad\;\,\quad\qquad$}
\end{figure}

We consider a downlink communication scenario of ISTN in Fig. \ref{fig:system model},
where $M$ TBSs and $N$ LEO satellites are deployed to provide service
to $J$ GUs. In addition, there are $L$ GEO satellite ground stations
(GEO-GSs) distributed in this area. The GEO satellite and LEO satellites
share the same frequency band to serve the GEO-GSs and TBSs, respectively.
The sets of GEO-GSs, TBSs and GUs are denoted by $\mathcal{L\textrm{=}}\left\{ 1,...,l,\ldots,L\right\} $,
$\mathcal{M\textrm{=}}\left\{ 1,...,m,\ldots,M\right\} $ and $\mathcal{J\textrm{=}}\left\{ 1,...,j,\ldots,J\right\} $,
while the set of LEO satellite is denoted as $\mathcal{N\textrm{=}}\left\{ 1,...,n,\ldots,N\right\} $.
In ISTN, TBSs can store popular content for data offloading, enabling
GUs to access the network through the available subchannels (SCs)
of their associated TBSs. LEO satellites provide backhaul service
to GUs through TBSs \cite{2021Joint}. Each TBS carries multiple independent
antennas to establish connections with multiple LEO satellites simultaneously
\cite{2019Ultra}. Let $\mathcal{T\textrm{=}}\left\{ 1,...,t,\ldots,T\right\} $
denote the set of timeslots (TSs). We assume that channel state information
(CSI) between LEO satellites and TBSs is known at all timeslots. 

We difine the C-band and Ka-band bandwidth as $\mathit{B}_{\textrm{C}}$
and $\mathit{B}_{\textrm{Ka}}$, respectively. To achieve higher transmission
rate, full frequency reuse is considered and the co-channel interference
cannot be ignored. The main types of interference in ISTN include
intra-system co-channel interference and inter-system collinear interference
between LEO and GEO satellites. 

\subsection{Caching Model\label{subsec:Caching Model}}

The local caching capability enables data offloading and traffic splitting,
which reduces network backhaul pressure \cite{9693912}. Through caching
files at TBSs, wireless caching enables GUs to download the requested
files directly from the TBSs' local storage, thereby reducing the
traffic overhead and communication time. Let $\mathcal{F}$ = $\left\{ 1,...,f,\ldots,F\right\} $
denote the set of requested files, which contains $F$ files. The
content popularity vector is denoted by $Q\textrm{=}\textrm{\ensuremath{\left\{  q_{f}\mid f=1,2,\ldots,F\right\} } }$,
where $q_{f}$ represents the popularity of the file $f$. The content
popularity $q_{n}$ follows the Zipf distribution \cite{2009Characteristics},
which is expressed as
\begin{equation}
q_{f}=\frac{f^{-\omega}}{\varOmega},\qquad\forall f\in\mathcal{F},\label{eq:request probability-1-1}
\end{equation}
where $\stackrel[f=1]{F}{\sum}q_{f}=1$ and $\omega$ is the Zipf
distribution index. Files with high popularity are always requested,
so popularity can also be used as the probability of GU request. To
represent the caching status of the TBSs, we difine a binary matrix
$G[g_{m,j}]_{M\times J}$, where $g_{m,j}=1$ indicates that TBS $m$
caches the file requested by GU $j$, and $g_{m,j}=0$ otherwise.
It is assumed that all TBSs have the same caching capacity. To efficiently
use the caching capability of the TBS, each GU is allowed to request
only one file from the associated TBS at one TS.

\subsection{Terrestrial Transmission Model\label{subsec:Terrestrial transmission-1}}

In the terrestrial network, each GU associates to the TBS that provides
the maximum average received power \cite{8013155}. Each TBS can serve
multiple GUs while each GU can be associated to only on TBS. We define
an indicator variable $a_{m,j}^{t}$ to describe the connection relationship
between GUs and TBSs, where $a_{m,j}^{t}=1$ represents that GU $j$
is served by TBS $m$ in TS $t$, and $a_{m,j}^{t}=0$ otherwise.
The set of available SCs of the TBS is denoted by $\mathcal{C\textrm{=}}\left\{ 1,...,c,\ldots,C\right\} $.
For GUs, we difine a binary matrix $X[x_{m,j,c}^{t}]_{M\times J\times C}$
to describe their groups, where $x_{m,j,c}=1$ represents that GU
$j$ is served by TBS $m$ over SC $c$ in TS $t$ and $x_{m,j,c}=0$,
otherwise. Let $P[p_{m,j,c}^{t}]_{M\times J\times C}$ denote the
transmit power from TBS $m$ to GU $j$ over SC $c$ in TS $t$. Thus,
the received signal of GU $j$ in TS $t$ is given by

\begin{equation}
y_{m,j}^{t}=\sum_{m\subset M}\sum_{j_{i}\subset J}\sqrt{p_{m,j_{i},c}^{t}}h_{m,j,c}u_{m,j_{i},c}+N_{m},
\end{equation}
where $h_{m,j,c}$ is the channel gain from TBS $m$ to GU $j$, $u_{m,j_{i},c}$
represents the transmit signal from TBS $m$ to GU $j_{i}$ over SC
$c$ and $N_{m}$ represents the additive white Gaussian Noise (AWGN)
at TBS $m$. Thus, the Signal-interference-noise Ratio (SINR) of a
GU $j$ is obtained by

\begin{equation}
\gamma_{m,j,c}^{t}=\frac{p_{m,j,c}^{t}x_{m,j,c}^{t}|h_{m,j,c}|^{2}}{\underbrace{\underset{m_{i}\neq m}{\sum}\underset{j_{i}\neq j}{\sum}p_{m_{i},j_{i},c}^{t}x_{m_{i},j_{i},c}^{t}|h_{m_{i},j,c}|^{2}}_{Intra-system\,co-channel\,interference}+\sigma_{n}^{2}},
\end{equation}
where $\sigma_{n}^{2}$ is the AWGN power. Therefore, the transmission
rate of GU $j$ from TBS $m$ on SC $c$ is calculated as
\begin{align}
R_{m,j,c}^{t}=B_{\textrm{C}}\log_{2}\left(1+\gamma_{m,j,c}^{t}\right).\label{eq:the data rate of users-1}
\end{align}

Note that GU's transmission rate is constrained by the backhaul traffic.
When the GUs request files from LEO satellites through associated
TBSs, it is assumed that each download will generate the same traffic,
denoted as $U_{\textrm{back}}$. Therefore, the achieveable transmission
rate of GU $j$ that needs to request files from the LEO satellite
is reformulated as
\begin{align}
\underset{\left(g_{m,j}=0\right)}{R_{m,j,c}^{t}}=\min\left(R_{m,j,c}^{t},U_{\textrm{back}}\right).\label{eq:the data rate of backhaul users-1}
\end{align}

\subsection{\textcolor{black}{LEO Backhaul} Transmission Model\label{subsec:Satellite-terrestrial transmission}}

We assume that fixed spatial positions for LEO satellites at each
TS. We denote the set of available SCs of the LEO satellites by $\mathcal{K\textrm{=}}\left\{ 1,...,k,\ldots,K\right\} $.
In ISTN, TBSs are under multiple satellite coverage, and each TBS
can be served by $N_{\mathit{\mathit{r}}}$ satellites simultaneously.
Due to the movement of LEO satellites, TBSs need to switch to other
satellites to ensure service continuity. A binary matrix $B[b_{n,m,k}^{t}]_{N\times M\times K}$
is provided to represent the relationship between LEO satellites and
TBSs, where $b_{n,m,k}^{t}=1$ indicates that TBS $m$ is served by
LEO satellite $n$ over SC $k$ in TS $t$. The elevation angle between
LEO satellite $n$ and TBS $m$ at TS $t$ can be expressed as $\theta_{n,m}^{t}$,
which can be calculated from the satellite position information predicted
by the TBSs. The transmission rate of GUs is limited by LEO-TBS backhaul
links. Specifically, the backhaul traffic consumed by GUs requesting
files cannot exceed the backhaul capacity of TBSs. Accordingly, the
backhaul SINR of TBS $m$ from LEO satellite $n$ is calculated as
\begin{equation}
\gamma_{n,m,k}^{t}=\frac{b_{n,m,k}^{t}p_{n,m,k}^{t}h_{n,m,k}}{\underbrace{\underset{n_{i}\neq n}{\sum}\underset{m_{i}\neq m}{\sum}b_{n_{i},m_{i},k}^{t}p_{n_{i},m_{i},k}^{t}h_{n_{i},m,k}}_{Inter-satellite\,co-channel\,interference}+\sigma_{n}^{2}},
\end{equation}
where $p_{n,m,k}^{t}$ is the transmit power allocated to TBS $m$
by LEO satellite $n$ over SC $k$ in TS $t$ and $h_{n,m,k}$ is
the channel gain from LEO satellite $n$ to TBS $m$. And the backhaul
rate of TBS $m$ from LEO satellite $n$ is expressed as

\begin{equation}
R_{n,m,k}^{t}=B_{\textrm{Ka}}\log_{2}\left(1+\gamma_{n,m,k}^{t}\right),\label{eq: the rate of TST-satellite link}
\end{equation}
 Thus, the backhaul capacity of TBS $m$ is obtained

\begin{equation}
C_{m}^{t}=\sum_{n\in\mathcal{N}}\sum_{k\in\mathcal{K}}R_{n,m,k}^{t}.
\end{equation}

The downlink from the LEO satellites can potentially cause interference
to GEO-GS \cite{2014Inline}. For the GEO-GS $l$, the interference
caused by LEO satellites at TS $t$ is given by 

\begin{equation}
I_{l}^{t}=\underset{n\in\mathcal{N}}{\sum}\underset{m\in\mathcal{M}}{\sum}b_{n,m,k}^{t}p_{n,m,k}^{t}h_{n,l},\label{eq:CINR}
\end{equation}
where $h_{n,l}$ is the channel gain from LEO satellite $n$ to the
GEO-GS $l$.

\section{Problem Formulation and Decomposition\label{sec:FORMULATION AND ALGORITHMS}}

This section models a system sum rate maximization problem that satisfies
the inter-system interference constraint. Then the original problem
is decomposed into two subproblems to be solved.

\subsection{Problem Formulation\label{subsec:Problem Formulation}}

In ISTN, a high transmission rate of GU is expected to be conducted.
However, intra- and inter-system interference may limit the LEO backhaul
capacity and affect the transmission rate of GUs \cite{8922609}.
Therefore, the goal of interference management is to maximize the
system sum rate while ensuring that each GEO-GS satisfies the interference
constraints. Accordingly, the intra- and inter-system interference
management problem is formulated as 
\begin{align}
\underset{\left\{ X,B,P\right\} }{\textrm{max}} & \sum_{m\in\mathcal{M}}\sum_{j\in\mathcal{J}}\sum_{c\in\mathcal{C}}R_{m,j,c}^{t}\label{eq:SRM}\\
s.t.\textrm{C1}: & \;x_{m,j,c}^{t}\leq a_{m,j}^{t},\forall m\in\mathcal{M},\forall j\in\mathcal{J},\forall t\in\mathcal{T},\tag{11a}\label{eq:a}\\
\textrm{C2}: & \sum_{m\in\mathcal{M}}\sum_{c\in\mathcal{C}}x_{m,j,c}^{t}\leq1,\forall j\in\mathcal{J},\forall t\in\mathcal{T},\tag{11b}\label{eq:b}\\
\textrm{C3}: & \sum_{j\in\mathcal{J}}x_{m,j,c}^{t}\leq1,\forall m\in\mathcal{M},\forall c\in\mathcal{C},\forall t\in\mathcal{T},\tag{11c}\label{eq:c}\\
\textrm{C4}: & \theta_{n,m}^{t}\geq\theta_{min},\forall n\in\mathcal{N},\forall m\in\mathcal{M},\forall t\in\mathcal{T},\tag{11d}\label{eq:d}\\
\mathit{\textrm{C5}:} & \sum_{n\in\mathcal{N}}\sum_{k\in\mathcal{K}}b_{n,m,k}^{t}\leq N_{r},\forall m\in\mathcal{M},\forall t\in\mathcal{T},\tag{11e}\label{eq:e}\\
\mathit{\textrm{C6}:} & \sum_{m\in\mathcal{M}}b_{n,m,k}^{t}\leq1,\forall n\in\mathcal{N},\forall k\in\mathcal{K},\forall t\in\mathcal{T},\tag{11f}\label{eq:f}\\
\mathit{\textrm{C7}:} & \!\sum_{j\in\mathcal{J}}\!\sum_{c\in\mathcal{C}}\!x_{m,j,c}^{t}\!\left(\!1\!-\!g_{m,j}\!\right)\!U_{\textrm{back}}\!\leq\!C_{m}^{t}\!,\forall m\in\mathcal{M}\!,\forall t\in\mathcal{T}\!,\tag{11g}\label{eq:g}\\
\textrm{C8}: & \;\sum_{j\in\mathcal{J}}\sum_{c\in\mathcal{C}}x_{m,j,c}^{t}p_{m,j,c}^{t}\leq P_{\textrm{TBS}},\forall m\in\mathcal{M},\forall t\in\mathcal{T},\tag{11h}\label{eq:h}\\
\textrm{C9}: & \;I_{l}^{t}\leq I_{th},\forall t\in\mathcal{T}.\tag{11i}\label{eq:i}
\end{align}
C1 indicates that each GU is associated with only one TBS, while C2
and C3 restrict the number of SCs assigned to each GU and the number
of GUs assigned to each SC, respectively. C4 constrains the selection
range of target LEO satellites for each TBS. C5 and C6 indicate that
each TBS can be served by $N_{r}$ satellites simultaneously. The
backhaul link constraint in ISTN is represented by C7, while C8 ensures
that the power allocated to each SC is within the maximum power of
the TBS. Finally, C9 denotes that the interference to GEO-GS should
meet the maximum interference threshold $I_{th}$ requirements.

The optimization problem \eqref{eq:SRM} is a non-convex integer nonlinear
programming problem with NP complexity. Due to the backhaul link constraint,
a direct transformation of problem \eqref{eq:SRM} into a convex optimization
problem is difficult. To address this, we adopt the Lagrangian dual
method to decompose the original problem \eqref{eq:SRM} into two
subproblems, enabling us to obtain the solution more efficiently.
Note that the result obtained in this way does not always satisfy
the backhaul link constraint C7, so we reduce GUs accessing the TBSs
to ensure that the consumed backhaul traffic does not exceed the backhaul
capacity.

\subsection{Problem Decomposition\label{subsec:decompositon}}

Due to the fact that the backhaul link constraint is considered in
the model, this optimization problem (\ref{eq:SRM}) is not convex.
Hence, we consider the dual problem of the original problem. By introducing
non-negative Lagrangian multiplier $\lambda$, the dual optimum of
the problem in (\ref{eq:SRM}) is expressed as
\begin{equation}
d^{*}=\min_{\lambda\geq0}d(\lambda)\triangleq\min_{\boldsymbol{\lambda}\geq0}\underset{\left\{ X,B,P\right\} }{\max}\mathcal{L}(X,B,P,\lambda),\label{eq: dual function}
\end{equation}
where $\lambda=\{\lambda_{1},\lambda_{2},...\lambda_{M}\}$. 

Thus, the Lagrange function is defined as
\begin{align}
\begin{array}{l}
\mathcal{L}(X,B,P,\lambda)\\
=\sum_{m\in\mathcal{M}}\sum_{j\in\mathcal{J}}\sum_{c\in\mathcal{C}}R_{m,j,c}\\
+\sum_{m\in\mathcal{M}}\lambda_{m}\left(C_{m}-\sum_{j\in\mathcal{J}}\sum_{c\in\mathcal{C}}x_{m,j,c}(1-g_{m,j})U_{\textrm{back}}\right)\\
=\sum_{m\in\mathcal{M}}\sum_{j\in\mathcal{J}}\sum_{c\in\mathcal{C}}R_{m,j,c}\\
-\sum_{m\in\mathcal{M}}\lambda_{m}\sum_{j\in\mathcal{J}}\sum_{c\in\mathcal{C}}x_{m,j,c}(1-g_{m,j})U_{\textrm{back}}\\
+\sum_{m\in\mathcal{M}}\lambda_{m}C_{m}.
\end{array}\label{eq:largrange}
\end{align}

\begin{proposition} 

Assume that $f(x)$ and $d(\lambda)$ are the primary problem and
dual problem. In addition, $f_{m}(x)$ are the is the relaxed constraints.
The Lagrange multiplier is denoted by $\lambda$. The optimal solutions
of the primary problem and dual problem are denoted by $f^{*}$ and
$d^{*}$, respectively. The $d^{*}$ obtained is an upper bound to
the optimal solution $f_{0}^{*}$.

\label{proposition: upper bound}

\end{proposition}

\textit{Proof: }The Lagrange function is expressed as
\begin{align}
\mathcal{L}(x,\lambda)=f(x)+\sum_{m\in\mathcal{M}}\lambda_{m}f_{m}(x),\label{eq:proof-Lagrange}
\end{align}
Let $\stackrel{\:\bullet}{x}$ be an arbitrary point within the feasible
domain $\mathcal{D}$ of the primary problem. The corresponding dual
function (\ref{eq: dual function}) is denoted by
\begin{equation}
\begin{aligned}d(\lambda) & =f(\overset{\,\bullet}{x})+\sum_{m\in\mathcal{M}}(\lambda_{m}f_{m}(\overset{\,\bullet}{x}))\\
 & \geq f(\overset{\,\bullet}{x}).
\end{aligned}
\label{eq:proof-Lagrange-1-1}
\end{equation}

We observe that (\ref{eq:proof-Lagrange-1-1}) holds for any points
$\mathcal{D}$. Accordingly, the dual optimum is an upper bound to
the optimal solution of primary problem, expressed as
\begin{equation}
d^{*}\geq f^{*},\label{eq:proof-Lagrange-1-1-1}
\end{equation}
where $f_{0}^{*}$ and $d^{*}$ are the optimal solutions of primary
problem and dual problem.\qed

Note that $d^{*}$ is an upper bound of primary problem (\ref{eq:SRM}),
as stated in Proposition 1. We observe that (\ref{eq:largrange})
is determined by $X$, $B$ and $P$. Thus, the primary problem is
decomposed into the following two subproblems.

The inter-satellite interference and handover management subproblem
is given by
\begin{equation}
\begin{aligned}\mathbf{P1}:\underset{\left\{ \boldsymbol{B}\right\} }{\max} & \sum_{m\in\mathcal{M}}\lambda_{m}C_{m}\\
s.t. & \textrm{ C4-C9}.
\end{aligned}
\label{eq:P1}
\end{equation}
The goal of P1 is to maximize the backhaul capacity through the management
of inter-satellite interference while satisfying the constraints of
inter-system interference. Considering that TBSs can connect to multiple
LEO satellites simultaneously, P1 can be transformed into a many-to-one
matching problem. It is noted that the inter-system interference constraint
C9, which helps to reduce the feasible region of resource allocation.

The terrestrial resource allocation subproblem is formulated as
\begin{equation}
\begin{aligned}\mathbf{\mathbf{P2}}:\begin{array}{l}
\underset{\left\{ \boldsymbol{X,P}\right\} }{\max}\end{array} & \sum_{m\in\mathcal{M}}\sum_{j\in\mathcal{J}}\sum_{c\in\mathcal{C}}R_{m,j,c}\\
 & -\sum_{m\in\mathcal{M}}\lambda_{m}\sum_{j\in\mathcal{J}}\sum_{c\in\mathcal{C}}x_{m,j,c}(1-g_{m,j})U_{\textrm{back}}\\
s.t. & \textrm{ C1-C3,C7}.
\end{aligned}
\label{eq:P2-1}
\end{equation}
P2 aims at maximizing the system sum rate under the backhaul link
constraint by coordinating intra-system interference. To achieve this
goal, the GUs are classified into backhaul-GUs and local-GUs according
to file request status and cache status. Specifically, when the file
requested by the GU is cached by its associated TBS, denoted by $g_{m,j}=1$,
the GU is classified as a local-GU, otherwise, as a backhaul-GU. We
find that P2 can be divided into two parts: subchannel allocation
and power allocation. SC allocation problem can be transformed into
a one-to-one matching problem, and then we find a feasible solution
for the power allocation by the water injection power algorithm.

\section{Coordinated Intra- and Inter-system Interference Management Method
Design\label{sec:ALGORITHMS design}}

In this section, we first specifically describe the interaction between
resource allocation and handover, and propose interference management
combined with inter-satellite handover algorithm to optimize the allocation
and handover selection in each TS. Subsequently, we designe a user
association and resource allocation algorithm to optimize the downlink
sum rate of the entire ISTN while satisfying satellite-terrestrial
constraint. Finally, we dintroduce a coordinated intra- and inter-system
method and analyze its convergence and complexity.

\subsection{Interference Management combined with Inter-Satellite Handover\label{subsec:matching game for satellite-terrestrial communication}}

As presented in (\ref{eq:P1}), the satellite interference and handover
management problem is compounded by the dense deployment of LEO satellites,
which results in inter-satellite interference, inter-system collinear
interference and frequent inter-satellite handover. Considering that
TBSs can be served by multiple LEO satellites, P1 can be reformulated
as a many-to-one matching problem subject to the inter-system constraint
\cite{2004Convex}. After the inter-system interference constraint
is satisfied through inter-satellite handover, multiple LEO satellites
are matched with a single TBS to improve the backhaul capacity. Specifically,
if the CINR of the GEO-GS is lower than the interference threshold
$I_{th}$, the inter-satellite handover is guided according to the
received signal strength, and the TBS will connect to the LEO satellite
with less interference to avoid inter-system interference.

1) \textit{Preparation} \textit{process}: Define $\varPhi$ is the
mapping of the set $\{\mathcal{M}\bigcup\mathcal{N}\bigcup\mathcal{K}\}$.
To jointly consider LEO satellites and SCs, we use $(l,k)$ to represent
the LEO-SC unit. The goal of a matching $\varPhi$ is to obtain a
stable $(m,(n,k))$ matching pair.

2) \textit{Preference relationship}: To eliminate TBS-satellite interference,
SIC is introduced in the matching process. Considering the order of
SIC decoding, the preference of TBS-satellite-SC unit $(m,(n,k))$
is denoted by
\begin{align}
\rho_{m,(n,k)}^{m',(n'_{'},k)}=\begin{cases}
|h_{n',m',k}|^{2}/|h_{n,m,k}|^{2}, & h_{n',m',k}>h_{n,m',k},\\
|h_{n',m',k}|^{2}, & h_{n',m',k}\leq h_{n,m',k}.
\end{cases}\label{eq:preference for satellite-subchannel unit}
\end{align}
If $\rho_{m,(s,k)}^{m_{1},(s_{1},k)}>\rho_{m,(s,k)}^{m_{2},(s_{2},k)}$,
the matched pair is more likely to be updated as $(m_{1},(s_{1},k))$
rather than $(m_{2},(s_{2},k))$. 

3) \textit{Matching process:} Initially, we sorted the TBS-satellite
units in descending order of preference and each SC is matched in
order. Specifically, to balance the quality and quantity of TBS-satellite
links, unmatched TBSs select the worst link satisfying
\begin{align}
m^{*},(n^{*},k)=\arg\max_{m\in\mathcal{M}_{un}}\min_{n\in\mathcal{N}_{un,k}}h_{n,m,k}.\label{eq:IGSSIC ini}
\end{align}
In (\ref{eq:IGSSIC ini}), $\mathcal{M}_{un}$ is the set of TBSs
connected to fewer than $N_{r}$ satellites. $\mathcal{N}_{un,k}$
is the set of unmatched LEO satellites over SC $k$.

Considering the co-channel interference between matching pairs, we
allow potential matching pairs to accept the proposal of matched unit
$(m,n)$, and candidate matching pairs are defined as follows

\begin{equation}
(m',n')=\arg\max_{m'\in\mathcal{M}_{un},n'\in\mathcal{N}_{un,k}}\rho_{m,(n,k)}^{m',(n',k)},\label{eq:SAT-subchannel 1}
\end{equation}
\begin{equation}
(m',n_{g}')=\arg\max_{m'\in\mathcal{M}_{un},n_{g}'\in\mathcal{N}_{un,k}^{g}}\rho_{m,(n,k)}^{m',(n_{g}',k)}.\label{eq:SAT-subchannel 2}
\end{equation}
In (\ref{eq:SAT-subchannel 1}) and (\ref{eq:SAT-subchannel 2}),
$\mathcal{N}_{un,k}^{g}$ is the set of LEO satellites whose channel
gain is greater than that of the matching pair $(m,(n,k)$), so that
co-channel interference between matching pairs can be avoided by introducing
SIC. The set $\mathcal{T}$ consists of candidate matching pairs $(m',(n',k)$)
and $(m',(n_{g}',k))$. 

4) \textit{Handover process:} The interference of GEO-GS is calculated
according to (\ref{eq:IGSSIC ini}). If the value is lower than $I_{th}$,
the interference caused by the LEO satellites in each visible range
is calculated according to the received signal power of GEO-GS. For
a TBS, when the difference between the received signal strength and
interference signal of other LEO satellites is higher than the currently
connected LEO satellite, it will choose the satellite with less interference
to connect. Therefore, we define a handover utility function, which
is represented by the difference between the received signal strength
of TBS and the interference signal strength of GEO-GS. The larger
the utility function, the greater the benefit of handover the TBS
to a new LEO satellite, which can reduce the interference to GEO satellites.
The handover utility is denoted by

\begin{equation}
U_{n}=\stackrel[m=1]{N_{\textrm{M}}}{\sum}\stackrel[k=1]{N_{\textrm{K}}}{\sum}b_{n,m,k}p_{n,m,k}h_{n,m,k}-\stackrel[m=1]{N_{\textrm{M}}}{\sum}\stackrel[k=1]{N_{\textrm{K}}}{\sum}b_{n,m,k}p_{n,m,k}h_{n,g}
\end{equation}

To reduce the handover frequency and increase the handover gain, a
threshold $H$ is set. When the handover utility of a new LEO satellite
$n$ exceeds the threshold $H$, the handover is triggered. After
the original satellite is switched to the new satellite, the new satellites
will provide backhaul services to TBSs to respond to user requests.
Then, update the candidates in $\mathcal{T}$ and proceed with further
matching. The utility of SC $k$ is set as the sum transmission rate
achieved by all TBSs over it. Then, each SC chooses a candidate matching
pair $(m'',(n'',k))$ based on the utility, expressed as
\begin{equation}
\mathcal{R}_{m}=\stackrel[m=1]{N_{\textrm{M}}}{\sum}\lambda_{m}C_{m}.
\end{equation}
Thus, the matching rules is defined as
\begin{align}
\varPhi_{1}\succ_{k}\varPhi_{2}\Leftrightarrow\mathcal{R}_{k}(\varPhi_{1})>\mathcal{R}_{k}(\varPhi_{2}).\label{eq:preference for satellite subchannel c}
\end{align}
If a TBS receives proposals from more than $N_{r}$ SCs, it sorts
these units in descending order of utility, and accepts the front
$N_{r}$ units. Then, other units are rejected.

5) \textit{End of the algorithm: }All SCs update the optimal matching
$\varPhi^{\ast}$ based on preference relations until convergence.

At initialization, (\ref{eq:IGSSIC ini}) is used to consider links
with average quality to reduce co-channel interference. Then, in the
matching process, if there exist LEO satellites that have already
been matched but cause excessive interference to GEO-GS, it is removed
from the matching set $\mathcal{T}$. The matching process will be
re-executed based on the utility to obtain the optimal matching $\varPhi^{\ast}$.

The algorithm for interference management combined with inter-satellite
handover (IMISH) is shown in Algorithm \ref{alg:Interference Management Algorithm}.
\begin{algorithm}[tbh]
\caption{Interference Management combined with Inter-Satellite Handover (IMISH)
\label{alg:Interference Management Algorithm}}

\small  

\textbf{Input: }Sets of TBSs, LEO satellites and SCs $\mathcal{M},$
$\mathcal{L}$ and $\mathcal{K}$; interference threshold of GEO-GS
$I_{th}$.

\textbf{Output: }The stable matching $\varPhi^{\ast}$.

\begin{algorithmic}[1]

\STATE \textbf{Initialization}

\STATE Record the current matching as $\varPhi$. Construct $\mathcal{M}_{un}=\mathcal{M}$
and $\mathcal{N}_{un,k}=\mathcal{N}$.

\WHILE{there exists a unmatched SC $k$}

\STATE Each unmatched SC $k$ makes a proposal to its most preferred
TBS-satellite unit $(m,n)$ according to (\ref{eq:IGSSIC ini}).

\IF {the unit $(m,n)$ receives proposals from more than $N_{r}$
SCs}

\STATE The $(m,n)$ accepts the first $N_{r}$ SCs in utility ranking
and rejects the others.

\ELSE 

\STATE The $(m,n)$ and $k$ form a new matching pair.

\ENDIF

\STATE The TBS $m$ is removed from $\mathcal{M}_{un}$.

\ENDWHILE\

\STATE\textbf{Matching Process}

\WHILE{ $\mathcal{M}_{un}$ is not $\varnothing$ or at least one
satellite-SC unit $(l,k)$ tries to propose}

\STATE Set $\mathcal{T}=\varnothing$.

\FOR{each SC $k$}

\FOR{each matched satellite-SC unit $(n,k)$}

\STATE The unit $(n,k)$ makes a proposal to its most preferred pair
$(m',(n',k))$ according to (\ref{eq:SAT-subchannel 1}) and (\ref{eq:SAT-subchannel 2}).

\STATE Add the candidate pairs to $\mathcal{T}$.

\STATE Calculate the interference of the GEO-GS according to (\ref{eq:CINR}).

\STATE\textbf{Handover Process}

\WHILE{$I\leq I{}_{th}$}

\STATE Rank the LEO satellites in order of interference strength
and remove the most interfering one.

\STATE Select the candidate pairs with the highest utility that meet
the handover threshold $H$. 

\STATE Update the candidate pairs in $\mathcal{T}$.

\ENDWHILE\

\ENDFOR\

\ENDFOR\

\FOR{each proposed TBS $m$}

\STATE Repeat step 6-8.

\STATE The matched LEO satellites are removed from $\mathcal{N}_{un,k}$.

\ENDFOR\

\ENDWHILE\

\RETURN the final matching $\varPhi^{\ast}$.

\end{algorithmic}
\end{algorithm}

\subsection{User Association and Resource Allocation\label{subsec:matching game for terrestrial}}

For P2, the backhaul capacity constraint is omitted to broaden the
scope of user association, thus ensuring a global optimization solution.
To coordinate intra-system co-channel interference, we formulate the
terrestrial resource allocation problem as a one-to-one matching.
Specifically, user association and SC allocation are performed based
on user distribution and channel conditions. After the optimal matching
is obtained, the downlink sum rate is improved through power allocation.
The algorithm is described as follows. 

1) \textit{Preparation} \textit{process}: Define $\eta$ is the mapping
between $\mathcal{J}$ and $\mathcal{M}\times\mathcal{C}$. We use
$(m,c)$ to represent the TBS-SC unit. The goal of a matching $\eta$
is to obtain a stable $(j,(m,c))$ matching pair that maximizes the
downlink sum rate.

2) \textit{Preference relationship}: There is a dependence among SCs
due to co-channel interference \cite{2019Ultra}. Specifically, the
transmission rate achieved on SC $c$ will be affected by other matched
SCs. Given a SC $c$, it expects to from a pair $(j,(m,c))$ with
a high channel gain GU $j$-TBS $m$ link and a low interference GU
$j$-TBS $m'$ link for higher transmission rate. Therefore, a preference
matrix $\Theta$ is defind to evaluate the effect on the TBS-SC unit
$(m,c)$ of all potential matching pairs as 
\begin{align}
\Theta_{j,(m,c)}^{j',(m',c)}=\left(h_{m',j',c}\right)^{\rho}/\left(h_{m,j',c}\right),\label{eq:TBS-subchannel preference}
\end{align}
where $\rho$ is the preference parameter. Then, if $\Theta_{j,(m,c)}^{j_{1},(m_{1},c)}>\Theta_{j,(m,c)}^{j_{2},(m_{2},c)}$,
the matched pair is more likely to be updated as $(j_{1},(m_{1},c))$
rather than $(j_{2},(m_{2},c))$. When $\rho=0$, the unit is concerned
with the impact of co-channel interference in matching process. When
$\rho=1$, the unit comprehensively considers the gain and co-channel
interference brought by new matching pairs.

3) \textit{Matching process:} Initially, each SC $c$ proposes to
its most preferred TBS $m$ and GU $j$, satisfying $a_{m,j}=1$.
This process can be formulated as
\begin{align}
(j^{*},m^{*})=\arg\max_{j\in\mathcal{J}_{un},0\leqslant m\leqslant N_{\textrm{M}}}h_{m,j,c},\label{eq:UDARA ini}
\end{align}
where $J_{un}$ is the set of unmatched GUs. 

According to the corresponding relationship between file requests
and TBSs cache, GUs are divided into local-UEs directly served by
TBSs and backhaul-GUs served by LEO satellites. Then, each TBS-SC
unit $(m,c)$ proposes to its most preferred matching pair, which
satisfies
\begin{equation}
(j',m_{l}')=\arg\max_{j_{l}'\in\mathscr{\mathcal{J}}_{un}^{l},m\in\mathcal{M}_{un,c}}\Theta_{j,(m,c)}^{j_{l}',(m',c)},\label{eq:TBS-subchannel local}
\end{equation}
\begin{equation}
(j',m_{b}')=\arg\max_{j_{b}'\in\mathcal{J}_{un}^{b},m\in\mathcal{M}_{un,c}}\Theta_{j,(m,c)}^{j_{b}',(m',c)}\text{,}\label{eq:TBS-subchannel back}
\end{equation}
where $\mathcal{J}_{un}^{l}$ and $\mathcal{J}_{un}^{b}$ represent
the sets of unmatched local-GUs and backhaul-GUs and $\mathcal{M}_{un,c}$
represents the set of unmatched TBSs over SC $k$. Sets of local and
backhaul candidate pairs $\mathcal{S}_{c}^{l}$ and $\mathcal{S}_{c}^{b}$
consisting of $(j_{l}',(m',c))$ and $(j_{b}',(m',c))$ are constructed
for each SC $k$, respectively. The utility of SC $k$ is defined
as

\begin{equation}
R_{c}^{l}=\stackrel[m=1]{N_{\textrm{M}}}{\sum}\stackrel[j=1]{N_{\textrm{J}}}{\sum}R_{m,j,c}.
\end{equation}

Then, each SC $k$ proposes to a candidate matching pair $(j_{l}^{''},(m^{''},c))$
in $\mathcal{S}_{c}^{l}$ with the highest utility. If the candidate
matching pair leads to a negative system gain, it will be rejected.
Instead, a new matching pair $(j_{b}^{''},(m^{''},c))$ is accepted
based on a modified utility function, which is given by
\begin{equation}
R_{c}^{b}=\stackrel[m=1]{N_{\textrm{M}}}{\sum}\stackrel[j=1]{N_{\textrm{J}}}{\sum}R_{m,j,c}-\stackrel[m=1]{N_{\textrm{M}}}{\sum}\lambda_{m}\stackrel[j=1]{N_{\textrm{J}}}{\sum}x_{m,j,c}\left(1-g_{m,j}\right)U_{\textrm{back}}.
\end{equation}
Accordingly, the matching rules is defined as
\begin{align}
\eta_{1}\succ_{c}\eta_{2}\Leftrightarrow R_{c}\left(\eta_{1}\right)>_{c}R_{c}\left(\eta_{2}\right),\label{eq:sunchannel preference}
\end{align}
where $R_{c}\left(\eta\right)$ represents the utility obtained by
SC $c$ under $\eta$. In addition, if GU $j$ receives multiple proposals,
it only accepts the one with the highest utility and others will be
rejected.

4) \textit{Power allocation}: After the SCs and GUs are matched, fine-grained
power allocation is implemented for each SC using the water injection
algorithm. When performing power allocation on the SCs in the TBS,
we consider the co-channel interference brought by other TBSs to remain
unchanged. Therefore, the power allocation of each TBS can be optimized
in a distributed manner through the water injection power algorithm.
The specific details of the algorithm are not discussed as the focus.
The power allocation aims at maximizing the system sum rate, achieved
by distributing the power proportionately to the channel gain. Specifically,
SCs with higher channel gain are allocated more power. The optimization
problem for power allocation is formulated as:

\begin{align}
\underset{\left\{ \boldsymbol{P}\right\} }{\max}\stackrel[m=1]{N_{\textrm{M}}}{\sum}\stackrel[j=1]{N_{\textrm{J}}}{\sum}\stackrel[c=1]{N_{\textrm{C}}}{\sum}\log_{2}\left(1+\frac{x_{m,j,c}p_{m,j,c}|h_{m,j,c}|^{2}}{I_{m,j,c}+\sigma_{n}^{2}}\right),\label{eq:water}
\end{align}
where $N_{\textrm{M}}$ denotes the GUs associated with TBS $m$ and
$I_{m,j,c}$ is the intra-system co-channel interference.

5) \textit{End of the algorithm: }All SCs update the optimal matching
pairs $\eta^{\ast}$ based on preference relations until convergence.

The algorithm for user association and resource allocation (UARA)
is shown in Algorithm \ref{alg:User Association Algorithm}. 
\begin{algorithm}[t]
\caption{User Association and Resource Allocation (UARA) \label{alg:User Association Algorithm}}

\small  

\textbf{Input: }Sets of TBSs, GUs and SCs $\mathcal{M}$, $\mathcal{J}$
and $\mathcal{C}$; association matrix $\boldsymbol{A}$.

\textbf{Output: }The stable matching $\eta^{\ast}$ and power allocation
matrix $P$.

\begin{algorithmic}[1]

\STATE\textbf{Initialization}

\STATE Record the current matching as $\eta$. Construct $\mathcal{M}_{un,c}=\mathcal{M}$,
$\mathcal{J}_{un}=\mathcal{J}$, $\mathcal{J}_{un}^{local}=\mathcal{J}^{local}$
and $\mathcal{J}_{un}^{back}=\mathcal{J}^{back}$.

\WHILE{there exists an unmatched SC $c$}

\STATE Each unmatched SC $c$ makes a proposal to the most preferred
GU-TBS unit $(j,m)$ according to (\ref{eq:UDARA ini}).

\IF {GU $j$ receives proposals from more than one SC}

\STATE The GU $j$ accepts the first unit in channel gain ranking
and rejects the others.

\ELSE 

\STATE The GU $j$ and $(m,c)$ form a new matching pair.

\ENDIF

\STATE The GU $j$ is removed from $J_{un}$.

\ENDWHILE\

\STATE\textbf{Matching Process}

\WHILE{$J_{un}$ is not $\varnothing$ or at least one TBS-SC unit
$(m,c)$ tries to propose}

\STATE Set $\mathcal{S}_{c}^{l}=\mathcal{S}_{c}^{b}=\varnothing$.

\FOR{each SC $c$}

\FOR{each matched TBS-SC unit $(m,c)$ }

\STATE The unit $(m,c)$ porposes to the most preferred pair $(j_{l}'',(m'',c))$
or $(j_{b}'',(m'',c))$ according to (\ref{eq:TBS-subchannel local})
and (\ref{eq:TBS-subchannel back}).

\ENDFOR\

\ENDFOR\

\FOR{each proposed GU $j$}

\STATE Repeat step 6-8. The matched local-GUs and backhaul-GUs are
removed from $\mathcal{J}_{un}^{l}$ and $\mathcal{J}_{un}^{b}$.

\STATE The matched TBS $m$ is removed from $\mathcal{M}_{un,c}$.

\ENDFOR\

\ENDWHILE\

\FOR{each matched UE-TBS-SC unit}

\STATE Update $P$ with $p_{m,j,c}$ by solving problem (\ref{eq:water})
with water injection power algorithm.

\ENDFOR\

\RETURN the final pair $\eta^{\ast}$ and power allocation matrix
$P$.

\end{algorithmic}
\end{algorithm}

\subsection{Coordinated Intra- and Inter-system Interference Management Algorithm}

The coordinated intra- and inter-system interference management (CIIM)
algorithm is designed to solve the intra- and inter-system interference
management problem. The proposed algorithm employs a many-to-one game-matching
approach involving SCs, TBSs, and LEO satellites, while also avoiding
collinear interference by inter-satellite handover. Additionally,
user association and SC allocation is conducted through one-to-one
matching of GUs, SCs and TBSs, and intra-system co-channel interference
is further managed through power allocation. The specific steps are
as follows: 
\begin{enumerate}
\item Given $\lambda^{(t)}$, satellite interference management and terrestrial
resource allocation subproblems are solved.
\item To satisfy the backhaul link constraint, if the backhaul traffic consumed
by the backhaul-GUs exceeds the backhaul capacity of the TBS, the
matching relationship will be removed according to transmission rate
of GUs \footnote{Note that the original problem is an integer programming problem and
the results obtained by the proposed method cannot always satisfy
the constraints \cite{2004Convex}.}.
\item Update the $\lambda$ by gradient descent formula, which is expressed
as $\lambda_{m}^{(t+1)}=\lambda_{m}^{(t)}-\theta^{(t)}\left(C_{m}-\stackrel[j=1]{N_{\textrm{J}}}{\sum}\stackrel[c=1]{N_{\textrm{C}}}{\sum}x_{m,j,c}(1-g_{m,j})U_{back}\right),\forall m\in\mathcal{M}$,
where $\theta^{(t)}$ denotes the monotonically decreasing exponential
function in the $t$-th iteration. The convergence criterion is $\mid\theta^{(t+1)}-\theta^{(t)}\mid\leq10^{-6}$.
Finally, the solution of intra- and inter-system interference management
problem is obtained. 
\end{enumerate}
The whole algorithm is shown in detail in Algorithm \ref{alg:JIMUA}.
\begin{algorithm}[tbh]
\caption{Coordinated Intra- and Inter-system Interference Management \label{alg:JIMUA}}

\small  

\begin{algorithmic}[1]

\STATE Initialize $X$, $B$, $P$, $\lambda$ and $\theta^{(t)}$.

\WHILE{$\mid\theta^{(t+1)}-\theta^{(t)}\mid\leq10^{-6}$}

\STATE Solve the satellite interference management problem and update
$B$ with $b_{l,m,k}$ by Algorithm 1.

\STATE Solve the terrestrial resource allocation problem and update
$X$ with $x_{m,j,c}$ and $P$ with $p_{m,j,c}$ by Algorithm 2.

\FOR{each TBS $m$}

\STATE The accessed backhaul-GUs are removed sequentially until (\ref{eq:g})
is satisfied.

\ENDFOR\

\STATE Update $\lambda$ and $\theta^{(t)}$.

\ENDWHILE\

\RETURN $X$, $B$ and $P$.

\end{algorithmic}
\end{algorithm}

\subsection{Convergence and Computational Complexity }

In this section, the convergence and computational complexity of the
proposed algorithms are analyzed. 
\begin{enumerate}
\item \textit{Convergence Analysis}. The convergence of algorithms mainly
depends on the matching process. The algorithm iteration stops when
a stable matching is reached. In each iteration, a preliminary match
is made, and then the matched units propose to the unmatched units
according to the preference relation. When there is no matching element
in the preference matrix of each matching element, it means that a
stable matching has been achieved, and then the iteration stops.
\item \textit{Computational Complexity} \textit{Analysis}. In IMISH algorithm,
only one TBS will be matched in each iteration when no handover occurs.
Thus, the number of iterations is $N_{\textrm{M}}$. If each SC accepts
a proposal in each iteration and $N_{\textrm{K}}$ matched units is
generated, the number of iterations is $\lceil N_{\textrm{M}}/N_{\textrm{K}}\rceil$.
Therefore, the iteration number is between $\lceil N_{\textrm{M}}/N_{\textrm{K}}\rceil$
and $N_{\textrm{M}}$. Futhermore, if the number of LEO satellites
connected to a TBS $m$ does not exceed $N_{\textrm{r}}$, each matched
TBS-satellite unit $(m,l)$ will construct a preference matrix with
a size of $N_{\textrm{M}}\times N_{\textrm{L}}$. Consequently, the
complexity is $O(N_{\textrm{M}}^{3}\times N_{\textrm{L}}\times N_{\textrm{K}})$
without handover. When the inter-satellite handover is required, the
worst case is that all LEO satellites need to be removed and the handover
utility of each LEO satellite needs to be calculated. Accordingly,
the complexity of IMISH algorithm is $O(N_{\textrm{M}}^{3}\times N_{\textrm{L}}^{2}\times N_{\textrm{K}})$.
Similarly, the iteration number of UARA algorithm is between $\lceil N_{\textrm{J}}/N_{\textrm{C}}\rceil$
and $N_{\textrm{J}}$. Then, power allocation needs to be implemented
on each SC $c$. Thus, the complexity of UARA is $O(N_{\textrm{T}}^{3}\times N_{\textrm{G}}\times N_{\textrm{C}}^{2}+N_{\textrm{T}}\times N_{\textrm{G}}\times N_{\textrm{C}})$.
Therefore, the complexity of CIIM is $O((N_{\textrm{M}}^{3}\times N_{\textrm{J}}\times N_{\textrm{C}}^{2}+N_{\textrm{M}}\times N_{\textrm{J}}\times N_{\textrm{C}})\times i_{max})$
considering $N_{\textrm{L}}\ll N_{\textrm{J}}$ in ISTN, where $i_{max}$
indicates the iteration numbers. 
\end{enumerate}

\section{Simulation Results \label{sec:Simulation Results Analysis}}

The simulation scene is a 3 km \texttimes{} 3 km area, where the GUs
and TBSs are subject to random distribution and uniform distribution
in the coverage of the satellites, respectively. The GEO and LEO satellites
provide service to TBSs, and the LEO satellite uses the actual trajectory
data from LEO satellites in the SpaceX constellations. The file caching
strategy of TBS adopts random scheme (RS) \cite{2018User}. And the
Zipf distribution index $\omega$ is set to 0.5. The total number
of the popular files in the terrestrial network is 50 and 40 files
are cached by the TBS. The terrestrial network adopts 5G communication
system, which has 273 resource blocks (RBs). And each LEO satellite
has 8 SCs. The total simulation time is 24 hours, divided into 1440
time slots. The channel between TBSs and GUs and the channel between
LEO satellites and TBSs are modeled as Rayleigh fading and Rician
fading, respectively. Other simulation parameters are set according
to \cite{2020Generalized}, as shown in Table \ref{tab:Major Simulation Parameters}. 

\begin{table}[t]
\centering{}\caption{\label{tab:Major Simulation Parameters}Simulation Parameters}
\begin{tabular}{|c|c|}
\hline 
\textbf{Parameter} & \textbf{Value}\tabularnewline
\hline 
\hline 
C-band carrier frequency $f_{\textrm{C}}$ & 4.9GHz\tabularnewline
\hline 
Ka-band carrier frequency $f_{\textrm{Ka}}$ & 30GHz\tabularnewline
\hline 
C-band carrier bandwidth $B_{\textrm{C}}$ & 100MHz\tabularnewline
\hline 
Ka-band carrier bandwidth $B_{\textrm{Ka}}$ & 500MHz\tabularnewline
\hline 
TBS transmit power & 47dBm\tabularnewline
\hline 
LEO Satellite transmit power & 48dBm\tabularnewline
\hline 
GEO Satellite transmit power & 60dBm\tabularnewline
\hline 
G/T parameter of each TBS and satellite & 18.5dB/K\tabularnewline
\hline 
Elevation angle of TBS & $30^{\circ}$\tabularnewline
\hline 
AWGN $\sigma_{n}^{2}$ & -174dBm/Hz\tabularnewline
\hline 
Preference parameters $\rho$ & 1\tabularnewline
\hline 
\end{tabular}
\end{table}

\subsection{Benchmark Algorithms}

In this section, we compare the proposed algorithm with existing algorithms
to evaluate the performance. For satellite interference management
and inter-satellite handover, five algorithms are given:
\begin{itemize}
\item \textbf{Joint interference management and user association (JIMUA)
algorithm: }As the description in \cite{9961951}, interference management
is addressed through collaborative computing. This method ignores
the impact of inter-satellite handover on interference management.
The complexity is $O(N_{\textrm{T}}^{3}\times N_{\textrm{G}}\times N_{\textrm{C}}\times i_{max})$.
\item \textbf{Minimum distance handover (MDH) algorithm:} In this algorithm,
minimum distance inter-satellite handover is introduced. The TBS will
switch from the LEO satellite connected in the current time slot to
the nearest LEO satellite.
\item \textbf{Random resource allocation and inter-satellite handover management
(RRAIHM) algorithm:} In this algorithm, LEO satellites randomly allocate
SCs to TBSs and other steps are similar to IMISH.
\item \textbf{Greedy algorithm: }In this algorithm, LEO satellite SCs are
allocated to TBSs, which can achieve the highest backhaul rate.
\item \textbf{Random algorithm: }In this algorithm, LEO satellite SCs are
randomly allocated to TBSs.
\end{itemize}
For terrestrial resource allocation, two algorithms are given:
\begin{itemize}
\item \textbf{Exhaustive search (ES) algorithm}: In ES algorithm, the optimal
solution is obtained by exhaustively enumerating all possible matching
combinations. The complexity is $O(N_{\textrm{J}}^{N_{\textrm{M}}\times N_{\textrm{C}}})$.
\item \textbf{User association and average allocation (UAAA) algorithm:
}In UAAA algorithm, TBSs equally distributes the power to each SCs
and other steps are similar to UARA. 
\end{itemize}

\subsection{Comparison with Benchmark Algorithms\label{subsec:Comparison-with-Benchmark}}

A small-scale test is conducted to evaluate the gap between UARA and
ES. Table \ref{tab:ES} shows the performance comparison of UARA and
ES for varying numbers of GUs and SCs. It is evident that the computation
time of ES increases rapidly as $N_{\textrm{J}}$ and $N_{\textrm{C}}$
increase. Meanwhile, when $N_{\textrm{J}}>N_{\textrm{M}}\times N_{\textrm{C}}$,
UARA can significantly save computation time while maintaining a <1\%
loss compared to ES.

\begin{table}[t]
\centering{}\caption{\label{tab:ES}Compared with ES}
\begin{tabular}{|c|c|c|c|}
\hline 
\multirow{2}{*}{$N_{\textrm{M}}$} & \multirow{2}{*}{$N_{\textrm{J}},N_{\textrm{C}}$} & \multicolumn{2}{c|}{Time (s), Rate (Mbps)}\tabularnewline
\cline{3-4} \cline{4-4} 
 &  & ES & UARA\tabularnewline
\hline 
\hline 
\multirow{5}{*}{2} & 4, 2 & 0.030797, 19.3615 & 0.000668, 19.3615\tabularnewline
\cline{2-4} \cline{3-4} \cline{4-4} 
 & 5, 2 & 0.262131, 21.2870 & 0.000737, 21.2870\tabularnewline
\cline{2-4} \cline{3-4} \cline{4-4} 
 & 6, 2 & 0.578923, 22.4408 & 0.001158, 22.4402\tabularnewline
\cline{2-4} \cline{3-4} \cline{4-4} 
 & 6, 3 & 76.30533, 32.1752 & 0.003353, 32.1752\tabularnewline
\cline{2-4} \cline{3-4} \cline{4-4} 
 & 7, 3 & 1187.87395, 33.7628 & 0.015913, 33.7609\tabularnewline
\hline 
\end{tabular}
\end{table}

\begin{figure}[t]
\centering{}\includegraphics[width=3.5in]{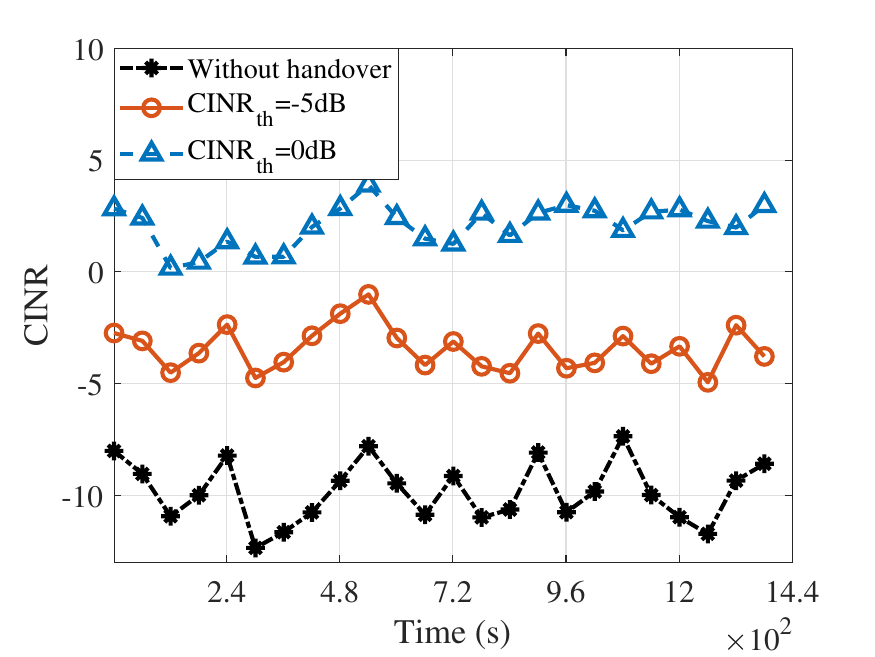} \caption{\label{fig:SNR 95} CINR of GEO-GS versus time under different working
thresholds.}
\end{figure}

\begin{figure}[t]
\centering{}\includegraphics[width=3.5in]{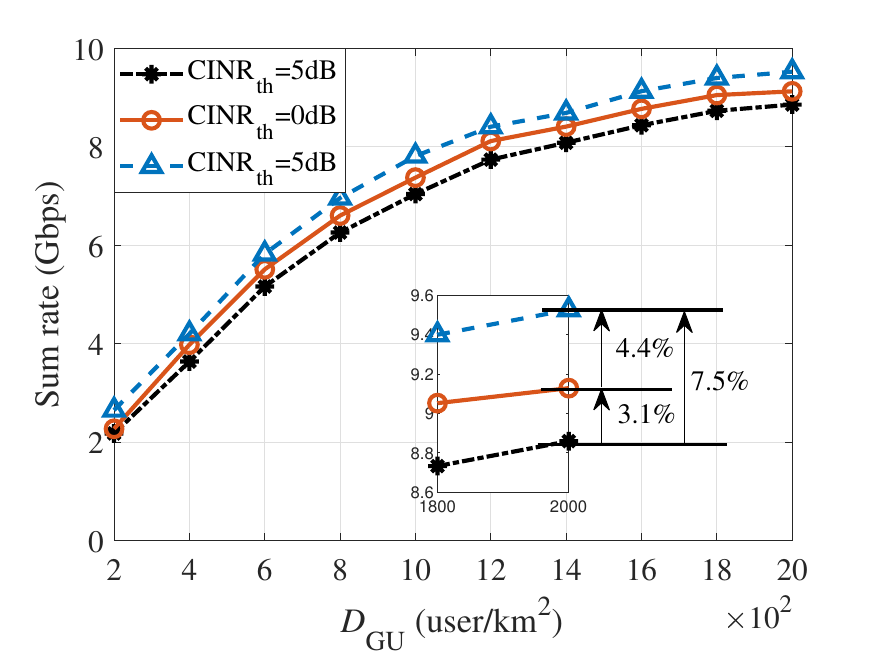} \caption{\label{fig:UDARA} Systen sum rate versus user density $D_{\textrm{GU}}$
under different $CINR_{th}$.}
\end{figure}

Fig. \ref{fig:SNR 95} illustrates the CINR of GEO-GS versus time
under different working thresholds. It can be found that inter-satellite
handover can effectively improve the CINR of the GEO-GS, leading to
an improvement of over 6.4 dB compared to the situation without handover.
The reason is that CIIM removes specific interference from LEO satellites
during the handover process. However, inter-satellite handover would
also degrade the backhaul capability, resulting in a decline in the
sum rate. Fig. \ref{fig:UDARA} shows how handover affects the system
sum rate. We can observe that as the working threshold increases,
the system sum rate will decrease. This can be attributed that as
the working threshold increases, the LEO satellites can be connected
by TBSs reduces. When user density $D_{\textrm{GU}}$ is high, excessive
handovers lead to a loss of 7.5\% of the sum rate. Accordingly, $CINR_{th}$=0dB
is adopted as a benchmark in the following simulations.

\begin{figure}[t]
\begin{centering}
\includegraphics[width=3.5in]{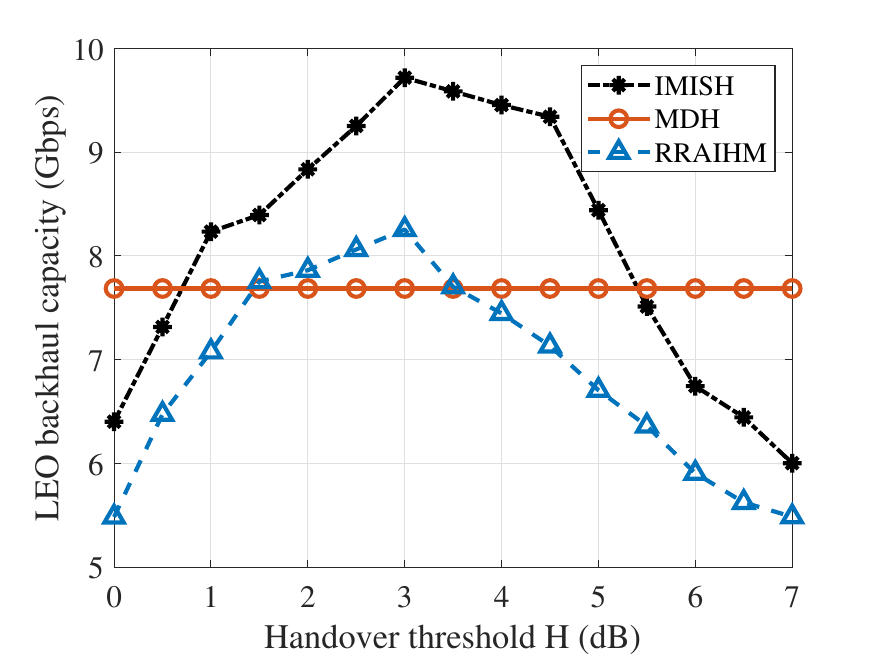}
\par\end{centering}
\centering{}\caption{\label{fig:SNR 95-1} LEO backhual capacity versus handover thresholds
under different schemes.}
\end{figure}

\begin{figure}[t]
\begin{centering}
\includegraphics[width=3.5in]{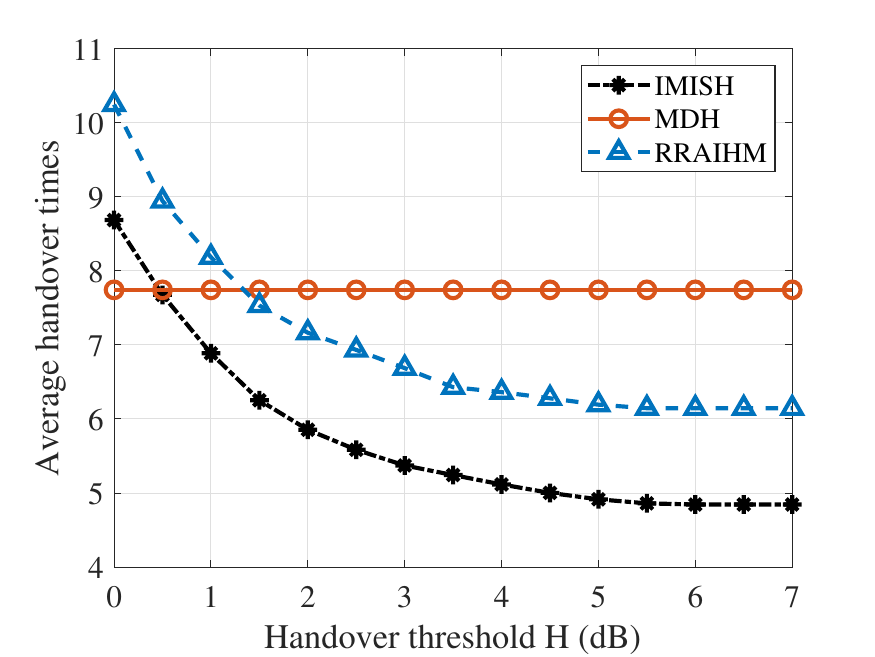}
\par\end{centering}
\centering{}\caption{\label{fig:SNR 95-1-1} Average handover times versus handover thresholds
under different schemes.}
\end{figure}

Figs. \ref{fig:SNR 95-1} and \ref{fig:SNR 95-1-1} depict the impact
of handover threshold $H$ on the LEO backhaul capacity and average
handover times. It can be seen that when the handover threshold $H$
increases, the backhaul capacity achieved by IMISH and RRAIHM will
first increase and then decrease. This is because a small handover
threshold is beneficial to handover triggering, resulting in a significant
increase in backhaul capacity. Especially, when the handover threshold
exceeds 3dB, the increase in handover threshold greatly limits the
improvement of backhaul capacity. Meanwhile, the increase in handover
threshold reduces the LEO satellites that meet the handover conditions,
which leads to a decrease in handover times. Since the handover criterion
adopted by MDH is the minimum distance, the change of handover threshold
does not affect its performance. In addition, the average handover
times of IMISH are always lower than other algorithms. The reason
is that during the matching process, the setting of the utility function
considers the influence of inter-satellite interference, so excessive
handover is not required. It is also noticed that compared with RRAIHM
and MDH, the backhaul capacity obtained by IMISH is increased by 17.7\%
and 29.8\%, respectively. The reason is that TBS selects LEO satellites
with higher gain and less interference, thereby reducing inter-satellite
interference. To balance the handover times and backhaul capacity,
$H$=3dB is used as a benchmark in subsequent simulations.

\begin{figure}[t]
\centering{}\includegraphics[width=3.5in]{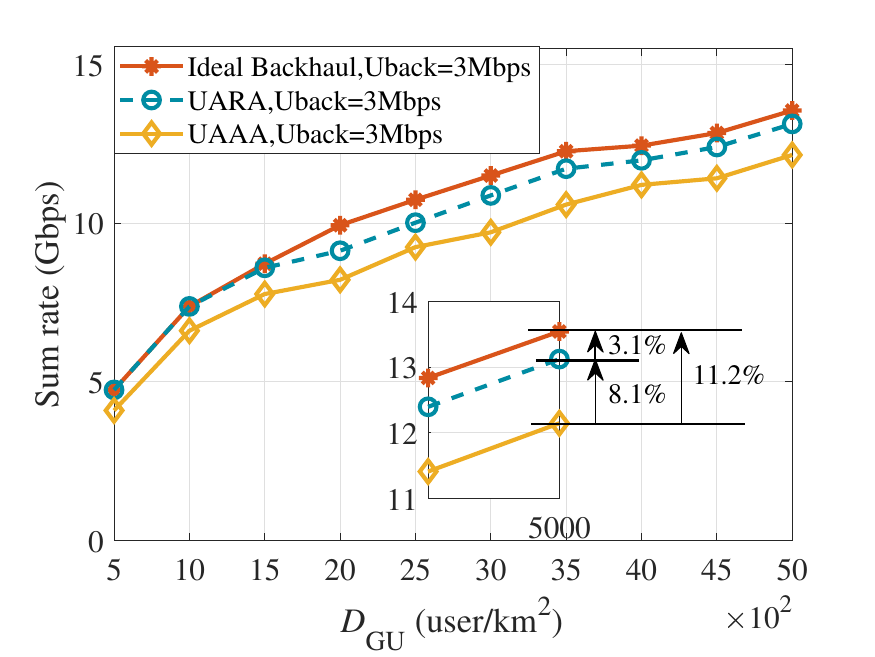}\caption{\label{fig:IGSSIC_rate} System sum rate versus user density $D_{\textrm{GU}}$
under different schemes ($U_{\textrm{back}}$=3Mbps).}
\end{figure}

In the context of combining UARA with the upper bound and UA, Fig.
\ref{fig:IGSSIC_rate} illustrates the performance comparison of downlink
sum rate as the user density increases under various algorithms. The
ideal backhaul is set for each TBS, so that the achieved sum rate
is considered as the upper bound. The results reveal that the sum
rate increases with $D_{\textrm{GU}}$. The reason is that the gain
brought by user access is more obvious than the impact of increased
interference. Compared with the upper bound, the lower backhaul capacity
achieved by CIIM results in a reduced sum rate. Compared with UA,
UARA performs power allocation according to user distribution and
channel conditions, thereby improving the sum rate. It is noteworthy
that the gap between the results obtained by UARA and the upper bound
is reduced as the user density increases, with only a 3.1\% loss at
$U_{\textrm{back}}$=3Mbps, which shows that UARA can effectively
increase the transmission rate through using decomposition and addressing
the original problem simultaneously.

\subsection{Performance of IMISH in Different Constellations}

\begin{figure}[t]
\centering{}\includegraphics[width=3.5in]{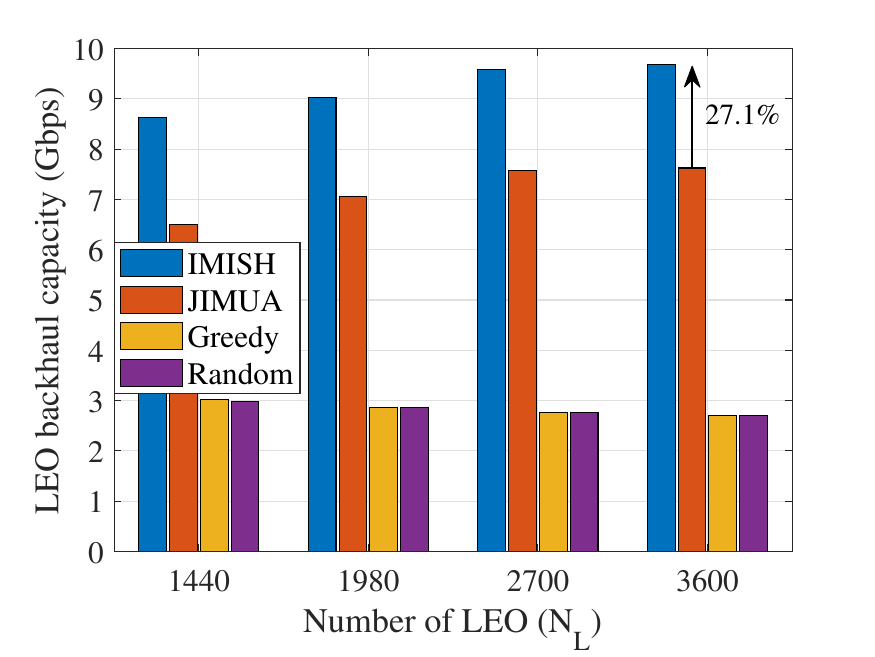} \caption{\label{fig:Nr} Comparison of LEO backhaul capacity with different
$N_{\textrm{r}}$ under different schemes ($H=3$dB and $CINR_{th}=0\textrm{dB}$).}
\end{figure}

We simulate the performance of IMISH under different $N_{\textrm{r}}$
and constellations to show its effectiveness for capacity improvement.
Fig. \ref{fig:Nr} simulates the backhaul capacity under different
algorithms when the number of connected satellites $N_{\textrm{r}}$
changes. As $N_{\textrm{r}}$ increases, the LEO backhaul capacity
achieved by different algorithms initially increases and then approaches
saturation. When $N_{\textrm{r}}$=3, the effect of the multi-connection
feature on the improvement of the LEO backhaul capacity reaches saturation.
This is because the co-channel interference caused by multiple connections
between LEO satellites and TBSs limits the gain, so that the obtained
backhaul capacity remains stable. 

\begin{figure}[t]
\centering{}\includegraphics[width=3.5in]{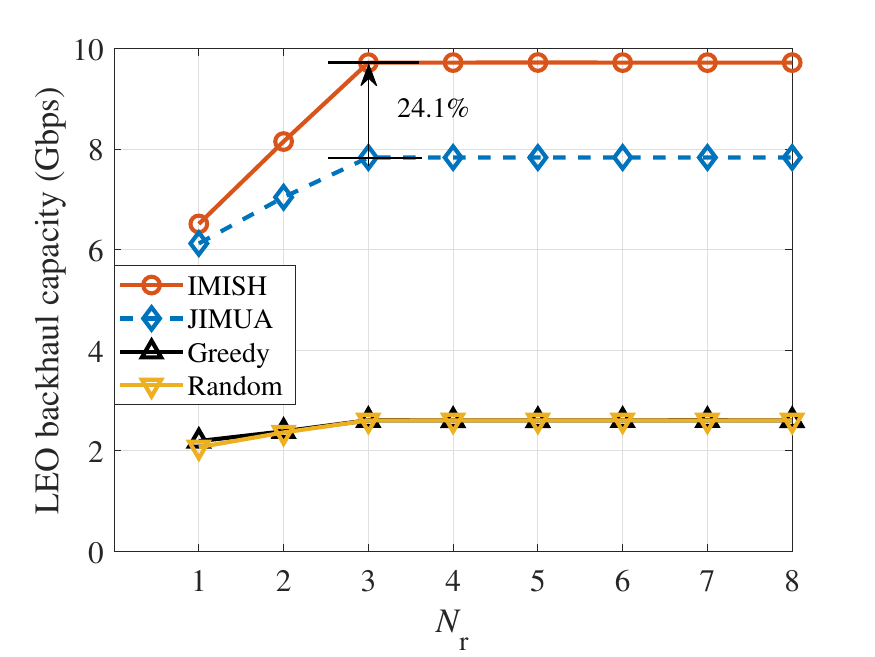} \caption{\label{fig:LEO} Comparison of LEO backhaul capacity with differen
$N_{\textrm{L}}$ under different schemes ($H=3$dB and $CINR_{th}=0\textrm{dB}$).}
\end{figure}

To further explore the situation of algorithm performance under different
numbers of satellites, we set up four satellite constellations. Each
type consists of 36 orbital planes with 40, 55, 75, and 100 LEO satellites
in each plane. Fig. \ref{fig:LEO} illustrates the backhaul capacity
obtained by different algorithms in different scale constellations.
It reveals that the backhaul capacity obtained by CIIM is always better
than that of JIMUA and keeps increasing as the size of the constellation
increases, while the other two algorithms are insensitive to the changes
in the number of satellites. This is because the IMISH considers the
impact of interference on link quality and improve the backhaul capacity
by inter-satellite handover. As the scale of the constellation continues
to increase, it is necessary to effectively manage the interference
within the constellation to increase the backhaul capacity.

\section{Conclusion\label{sec:Conclusion}}

In this paper, we investigate the coordinated design of intra- and
inter-system interference management and interference avoidance through
inter-satellite handover in ISTN. In particular, a coordinated intra-
and inter-system interference management (CIIM) scheme is proposed
to coordinate intra-system co-channel interference while satisfying
the constraint of inter-system interference. Through inter-satellite
handover, TBS can make handover decisions to avoid collinear interference
between GEO and LEO satellites. In addition, we propose a user association
and resource allocation scheme to conduct the user association and
coordinate the co-channel interference among GUs. The simulation results
show the effectiveness of CIIM in mitigating inter-system collinear
interference and improving the downlink sum rate of ISTN. Hence, this
study can offer the reference for overcoming the intra- and inter-system
interference in ISTNs.

\bibliographystyle{IEEEtran}
\phantomsection\addcontentsline{toc}{section}{\refname}\bibliography{bib}

\end{document}